\documentclass{article}

\usepackage{graphicx}
\usepackage{amssymb}

\usepackage{lineno}

\usepackage{subfig}
\usepackage{acronym}
\usepackage{amsfonts} 
\usepackage[hidelinks,bookmarks=false]{hyperref}
\usepackage{booktabs}
\usepackage{picture,xcolor}
\usepackage{threeparttable}
\usepackage{balance}
\usepackage{color}
\usepackage{caption}
\usepackage{multirow}
\usepackage{threeparttable}
\usepackage{array}

\usepackage{xcolor}

\begin{document}

\title{\textcolor{black}{\Large A Review on Communication Protocols for \Large Autonomous Unmanned Aerial Vehicles for \Large Inspection Application \\ }}

\author{\normalsize Liping Shi, Néstor J. Hernández Marcano, and  \normalsize Rune Hylsberg Jacobsen}

\date{\textcolor{black}{\emph{\small Department of Electrical and Computer Engineering}, \\ \emph{\small Aarhus University, Denmark} \\
\vspace{5 mm} PREPRINT }
}

\acrodef{3GPP}{Third Generation Partnership Project}
\acrodef{A2A}{Air-to-Air}
\acrodef{A2G}{Air-to-Ground}
\acrodef{AODV}{Ad Hoc On-Demand Distance Vector}
\acrodef{API}{Application Programming Interface}
\acrodef{BVLOS}{Beyond Visual Line of Sight}
\acrodef{C2}{Command and Control}
\acrodef{CSS}{Chirp Spread Spectrum}
\acrodef{DSSS}{Direct Sequence Spread Spectrum}
\acrodef{D2D}{Device-to-Device}
\acrodef{EASA}{European Union Aviation Safety Agency}
\acrodef{EC}{European Commission}
\acrodef{eMTC}{Enhanced Machine Type Communications}
\acrodef{G2A}{Ground-to-Air}
\acrodef{GSM}{Global System Mobile}
\acrodef{HWMP}{Hybrid Wireless Mesh Protocol}
\acrodef{IoT}{Internet of Things}
\acrodef{KPI}{Key Performance Indicator}
\acrodef{LPWAN}{Low Power Wide Area Network}
\acrodef{LTE}{Long Term Evolution}
\acrodef{LTE-A}{Long Term Evolution - Advanced}
\acrodef{LoRa}{Long Range}
\acrodef{LoRaWAN}{Long Range Wireless Area Networks}
\acrodef{MAC}{Medium Access Control}
\acrodef{MANET}{Mobile Ad Hoc Network}
\acrodef{MLME}{MAC Sublayer Management Entity}
\acrodef{MPM}{Mesh Peering Management}
\acrodef{MIMO}{Multiple Input Multiple Output}
\acrodef{M2M}{Machine-to-Machine}
\acrodef{NB-IoT}{Narrowband IoT}
\acrodef{OFDM}{Orthogonal Frequency Division Multiplexing}
\acrodef{PER}{Packet Error Rate}
\acrodef{PHY}{Physical Layer}
\acrodef{PSR}{Packet Success Rate}
\acrodef{ProSe}{Proximity-based Services}
\acrodef{RSRQ}{Reference Signal Received Quality}
\acrodef{ROS}{Robot Operating System}
\acrodef{ROS2}{Robot Operating System 2}
\acrodef{SAR}{Search and Rescue}
\acrodef{SME}{Station Management Entity}
\acrodef{UAS}{Unmanned Aircraft System}
\acrodef{UAV}{Unmanned Aerial Vehicle}
\acrodef{UE}{User Equipment}
\acrodef{UWB}{Ultra Wide-Band}
\acrodef{VLOS}{Visual Line of Sight}
\acrodef{V2X}{Vehicular-to-X}
\acrodef{WAVE}{Wireless Access in Vehicular Environments}
\acrodef{WiFi}{Wireless Fidelity}
\acrodef{WANET}{Wireless Ad Hoc Network}
\acrodef{WLAN}{Wireless Local Area Network}
\acrodef{WPAN}{Wireless Personal Area Network}
\acrodef{zeroconf}{Zero-configuration networking}
\acrodef{UDP}{User Datagram Protocol}
\acrodef{DDS}{Data Distribution Service}
\acrodef{RTPS}{Real-Time Publisher-Subscriber protocol}
\acrodef{QoS}{Quality of Service}

\maketitle

\begin{abstract}
\textcolor{black}{The communication system is a critical part of the system design for the autonomous \ac{UAV}. It has to address different considerations, including efficiency, reliability and mobility of the UAV.}
In addition, a multi-UAV system requires a communication system to assist information sharing, task allocation and collaboration in a team of UAVs. In this paper, we review communication solutions for supporting a team of UAVs while considering an application in the power line inspection industry.
\textcolor{black}{We provide a review of candidate wireless communication} technologies {for supporting communication in UAV applications.} 
Performance measurements and UAV-related channel modeling of those candidate technologies are reviewed. {A discussion of current technologies for building UAV mesh networks is presented. We \textcolor{black}{then analyze} the structure, interface and performance of robotic communication middleware, ROS and ROS2. Based on our review, the features and dependencies of candidate solutions in each layer of the communication system are presented.}

\end{abstract}

\subsubsection*{Keywords}
Wireless communication, mesh network, {robot operating system \textcolor{black}{(ROS)}}, Unmanned Aerial Vehicle (UAV).


\section{Introduction \label{sec:intro}}
{More and more industries start to consider involving autonomous \acp{UAV} to be part of their services or solutions. The first group of applications that start getting benefits from this new technology includes infrastructure}  inspections~\cite{Erdelj2017,Mansouri2017,Eschmann2012,Nikolic2013}, surveillance and reconnaissance~\cite{Acevedo2013,Berger2010}, \ac{SAR} {services}~\cite{Bernard2011, Bahnemann, Waharte2009}, as well as {autonomous picking and delivery}~\cite{Bernard2011,Nieuwenhuisen2017}.
In particular, the applications of rotor-winged \acp{UAV} for inspection operations have {drawn noticeable} attention from researchers and industrial professionals due to its large cost-saving potential~\cite{Deloitte, Bobbe2020, Pahwa2019, Mansouri2018}.

{Most use cases of today are conducted by controlling} a single \ac{UAV} \cite{Nikolic2013, Christiansen2017}. More recently, there is increasing {interest in the usage of multiple \acp{UAV} in order to improve the efficiency by collaboration}~\cite{Berger2010, Bahnemann, Waharte2009}. Although \ac{BVLOS} capable \acp{UAV} are appearing in the marketplace {from companies such as} Xcel Energy \cite{xcelenergy}, Delair \cite{delairdrone}, Microdrones \cite{microdrones}, PrecisionHawk \cite{precisionhawk} and Terra Drone \cite{terradrone}; \textcolor{black}{there is only a reduced amount of studies in the State-of-the-Art} that address the challenges of coordination of \acp{UAV} in \ac{BVLOS} applications.

\begin{figure}[t]
  \centering
  \includegraphics[width=\columnwidth]{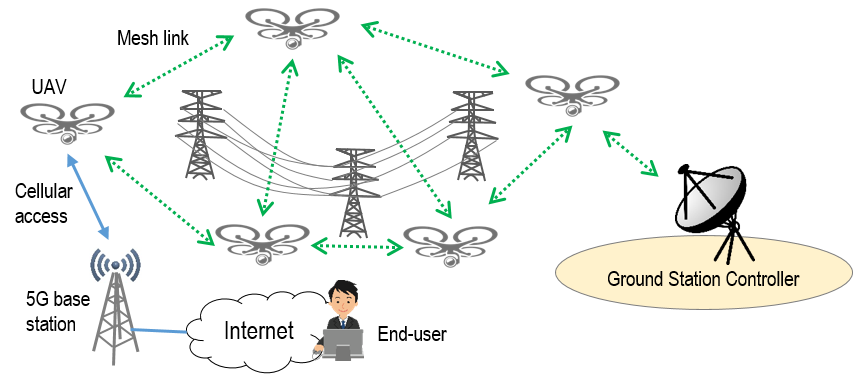}
  \caption{{Overall system concept of a \ac{UAV} power line inspection. A set of \acp{UAV} are interconnected through a wireless mesh network controlled locally from the ground. The \acp{UAV} also have access to cellular infrastructure for mission data offloading}.
  \label{fig:system-concept}}
\end{figure}

To ensure a high {level of uniform safety} of the European airspace, the \ac{EASA} has {published regulations} that contribute to the development of a common market while ensuring safe operations {in a private manner}.
The regulation follows the ``U-Space'' ambition of the \ac{EC} to ``make drone use in low-level airspace safe, secure and environmentally friendly''~\cite{Geister2019}.
The U-space {supports safe and secure UAV operations by providing} a wide range of services based on agreed EU standards.
{Lately}, \ac{EASA} approved the \ac{EC}'s proposal to regulate the operations and registration of \acp{UAS} in Europe. 
Essentially, \acp{UAV} must comply with a specific category of the new regulatory framework covering the regulation of all civil \acp{UAV} regardless of their maximum take-off masses~\cite{opinion01}.
{Following the regulation}, the communication system of the autonomous \ac{UAV} system {is required to} implement a reliable \ac{C2} connection flow to all \acp{UAV} of the cluster {with additional fail-safe mechanisms to deal with different communication issues}.

{Concerning the specification of UAV communication requirements, Andre et al. \cite{Andre2014} discussed the communication requirements raised by applications in micro aerial vehicle networks.
The work proposed a  hierarchical system representation based on building blocks and key functionalities to characterize communication needs.
Vinogradov et al. \cite{Vinogradov2018} discussed the requirements from the role of UAVs (e.g. aerial user, aerial base station) and their communication channel and scenarios.  Wang et al.~\cite{Wang2016} surveyed the requirements from the view of a UAV network structure and protocol architecture and discussed open challenges in the design for flying networks. Maruyama et al.~\cite{Maruyama2016} evaluated the performance of ROS and ROS2 communication middleware on different aspects.}


\textcolor{black}{Although the previous works consider different aspects related to the requirements, there is a lack of studies about unified reviews on the full communication protocol stack for UAVs. }To clarify potential solutions for the communication system layer by layer, a survey has been made \textcolor{black}{in~\cite{shi2019} discussing initially protocols for the physical and data link layers.}
\textcolor{black}{In this paper, we extend such survey to further access technologies but also protocols for the highers layers providing a holistic view. Our major contributions are as follows:}
{
\begin{itemize}
    \item We classify the reviewed literature with regards to their \ac{UAV} communication link type and wireless access technologies based on reported in-field tests.
    \item We provide a survey on studies regarding wireless mesh \textcolor{black}{networking protocols} targeted for \ac{UAV} applications.
    \item We analyze different robotic middlewares from a communication point of view, \textcolor{black}{evaluate their potential benefits}, architectures and \acp{API}.
\end{itemize}
}


{The paper is organized as follows: Section~\ref{sec:related} elaborates the targeted application that \textcolor{black}{a} communication system should be designed for. }Section~\ref{sec:wireless} presents studies of wireless technologies for multi-\ac{UAV} networks, where in-field measurements are reported to assess their performance. {Section~\ref{sec:meshnetworks} discusses surveyed wireless mesh network protocols aimed for the proposed application. At last, Section~\ref{sec:rosnetworks} introduces the robotic communication middleware and a comparison between ROS and ROS2. Conclusions are drawn in Section~\ref{sec:conclusion}.}


%
\section{Inspection Scenario \label{sec:related}}
{In this section, the project Drones4Energy is presented as an example to illustrate the target application. }
Drones4Energy aims to build a cooperative, autonomous and continuously operating UAV system that will be offered to power line operators to inspect the power grid frequently and autonomously \cite{ourproject}. The {proposed autonomous multi-UAV} system consists of UAVs that fly in a team along the cables to inspect {different facets} of the cables {and pylons. The motivation to implement a team of UAVs is to improve the inspection efficiency, i.e., extend total inspection range to compensate the influence of the short flight time due to limited battery capacity and the energy-intensive propulsion system.} 
{The system development consists of four sub-tasks, \textcolor{black}{namely:} autonomous flight, communication and coordination, inspection data analysis and autonomous energy harvest.  Autonomous flight~\cite{Ebeid2018} aims to support the UAV on perception, navigation and control in an outdoor environment with \textcolor{black}{high-resolution} RGB  cameras. Communication and coordination provide connectivity and network solutions to enable collaborative motion planning, flight formation, coordinated task allocation\textcolor{black}{,} etc. Inspection data analysis gives the system the ability to automatically process the inspection images and generate an analysis report.  The target of the energy harvest is to develop a loadable device for UAV to harvest energy on the powerline and its related flight control solution.}

\begin{figure}[th]
  \centering
  \includegraphics[scale = 0.9]{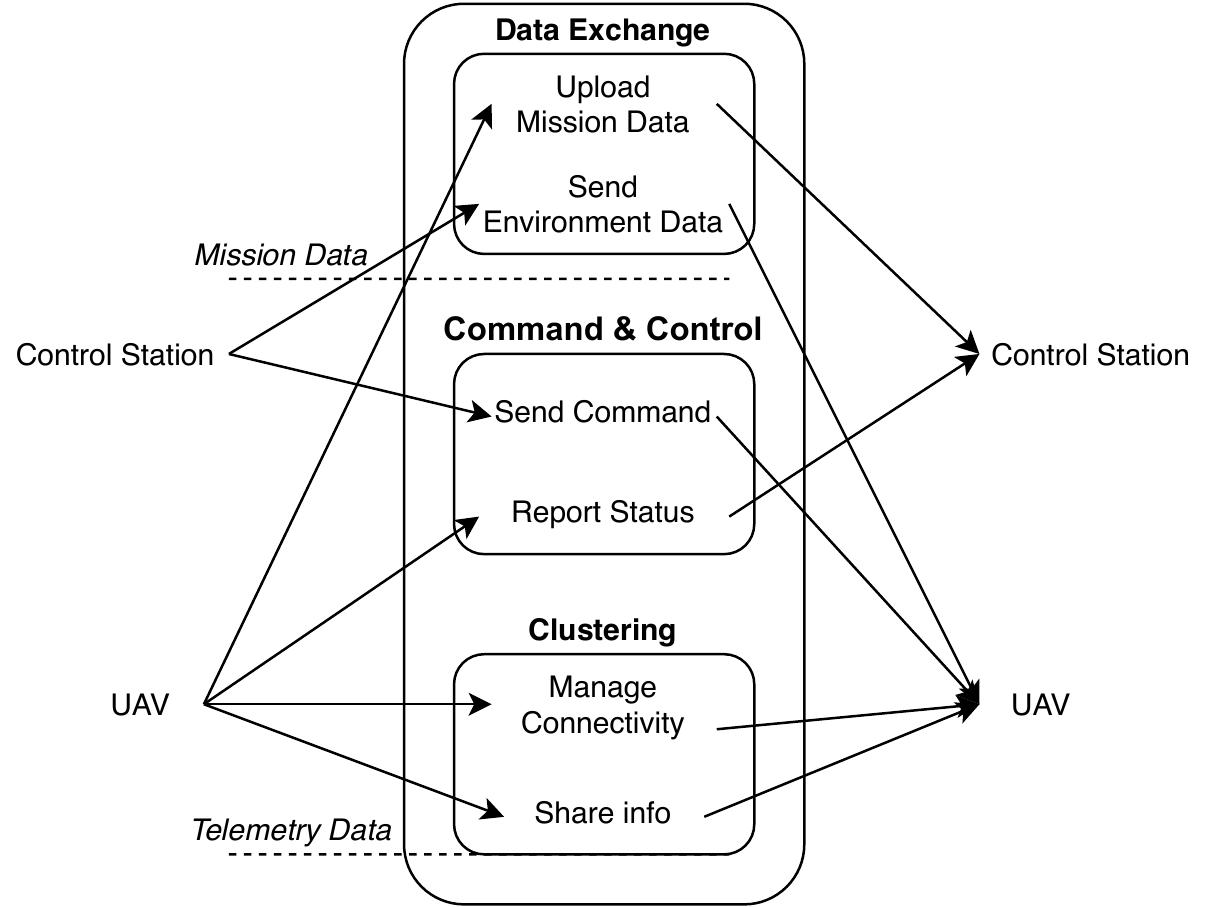}
  \caption{Use case diagram. 
Actors are \textbf{UAV}: A UAV within a cluster. \textbf{Control Station}: The control station \textcolor{black}{is designed} for providing access to the UAV system, high-scale computing/processing and storing of data. }
  \label{fig:use_case}
\end{figure}

\subsection{Use Case Specification}
{Fig.~\ref{fig:use_case} decomposes our inspection application using the autonomous multi-UAV system from the perspective of communication to a set of use cases which can be classified into three categories:} (i) Command and Control, (ii) Clustering, and (iii) Data Exchange. 

\textbf{Command and Control (C2)} are orchestrated from the mission control facility via the {Control} Station. Use cases in this category provide {data related to the} ongoing mission including reporting from the cluster and {establish} a channel to {transport} updates of mission commands to the UAVs {in the field}. Some (or all) UAVs in the system {behave} \textbf{Clustering (C)} by establishing a meshed wireless network. {It enables} inter-UAV communication {during the collaboration for executing} the high-level mission plan {received from the Control Station}. {At last,} the \textbf{Data Exchange (DE)} is carried in two different ways. {On one hand,} it is given by the transmission of mission data to the {Control Station. On the other hand,} it allows context {data} such as maps and/or weather forecasts to be {transported} to the UAVs.  {Two types of data \textcolor{black}{profiles} are recognized in the proposed system.} \textit{Mission data} is data generated from the payloads such as camera images {which often \textcolor{black}{have} comparatively large size and simple data structure}. \textit{Telemetry data} describes operational aspects of the missions such as {waypoints}, UAV {flight} status and communication indicators, {which is usually packed as small size messages but with \textcolor{black}{a} diversity of data structures}.

\begin{table}[]
\centering
\scriptsize
\caption{ {Communication types and characteristics.}}
\begin{tabular}{|l|l|l|l|l|l|}
\hline
\multicolumn{1}{|c|}{\textbf{Category}}                                        & \multicolumn{1}{c|}{\textbf{Use Case}}                               & \multicolumn{1}{c|}{\textbf{Flow Type}}                              & \textbf{Priority} & \textbf{\begin{tabular}[c]{@{}l@{}}Data Size\\ (Kbyte)\end{tabular}} & \textbf{\begin{tabular}[c]{@{}l@{}}Mobility\\(m/s)\end{tabular}}                                               \\ \hline
\multirow{2}{*}{\begin{tabular}[c]{@{}l@{}}Data \\ Exchange\end{tabular}}      & \begin{tabular}[c]{@{}l@{}}Upload\\ Mission Data\end{tabular}     & Unicast                                                         & Low               & $>1000$                                                           & \begin{tabular}[c]{@{}l@{}}Static or \\ $<1$\end{tabular} \\ \cline{2-6} 
                                                                               & \begin{tabular}[c]{@{}l@{}}Send\\ Environment\\ Data\end{tabular} & Unicast                                                         & Low               & $>1000$                                                          & \begin{tabular}[c]{@{}l@{}}Static or \\ $<1$\end{tabular} \\ \hline
\multirow{2}{*}{\begin{tabular}[c]{@{}l@{}}Command \\ \& Control\end{tabular}} & \begin{tabular}[c]{@{}l@{}}Send \\ Command\end{tabular}           & Unicast                                                         & High              & $<0.1$                                                           & $1 - 10$       \\ \cline{2-6} 
                                                                               & Report Status                                                     & Unicast                                                         & Low               & $<0.1$                                                          &  $1 - 10$        \\ \hline
\multirow{2}{*}{Clustering}                                                    & \begin{tabular}[c]{@{}l@{}}Manage \\ Connectivity\end{tabular}    & \begin{tabular}[c]{@{}l@{}}Multicast \\ \& Unicast\end{tabular} & Middle            & $<0.1$                                                           &  $1 - 10$        \\ \cline{2-6} 
                                                                               & Share Info                                                        & Unicast                                                         & Low               & \begin{tabular}[c]{@{}l@{}}$<100$ \end{tabular}      &  $1 - 10$        \\ \hline
\end{tabular}
\label{Tab:message_profile}
\end{table}
{
To clarify the communication characteristics for each category, Table~\ref{Tab:message_profile} is presented. We specify the communication in terms of flow type, priority, message size, and mobility. For the flow type, multicast is required for the case "Manage Connectivity" to support node discovery in the network. Unicast is widely preferred for other cases by concerning the limited throughput of the network. \textcolor{black}{To maintain the accessibility} of some important messages under busy network \textcolor{black}{conditions} and \textcolor{black}{design-related} flow control strategies, priority and data size are identified. While estimating the data size, We consider inspection images as typical mission data. Environment data includes flight related weather information and map data. Data sizes for command and status data are estimated based on the MAVLink common message set~\cite{mavlink}. In addition, mobility needs to be considered when addressing the problem of routing for maintaining end-to-end connectivity. The ground speed of the UAV is used to represent mobility.
}

\begin{figure}[th]
  \centering
  \includegraphics[scale = 0.9]{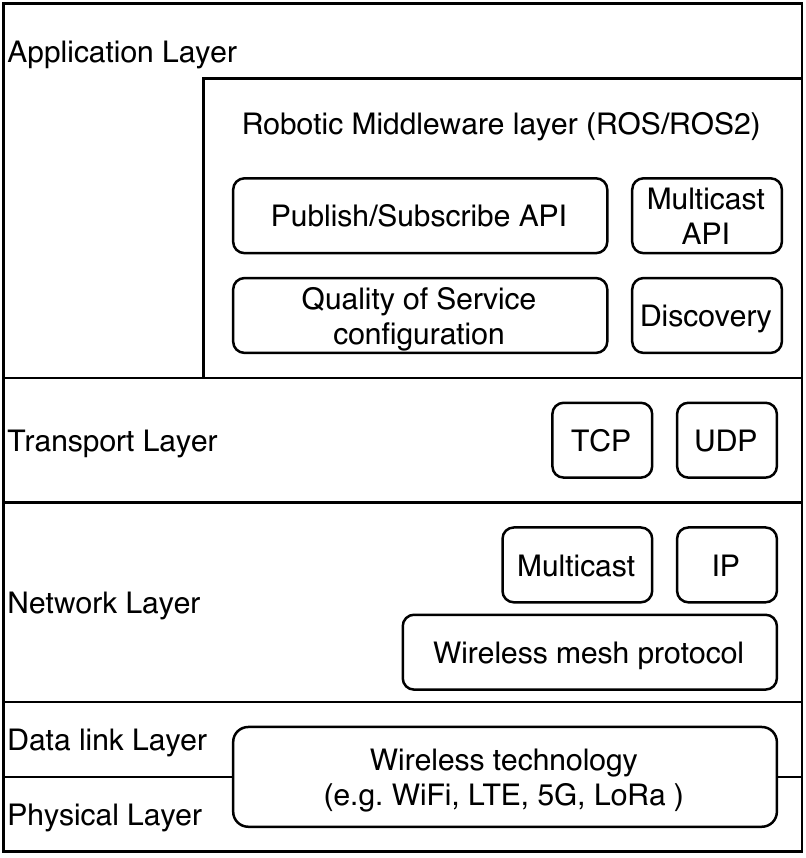}
  \caption{Protocol stack for the multi-UAV system}
  \label{fig:protocol}
\end{figure}
\subsection{Protocol stack}
{The multi-UAV communication system is formed with multiple layers with different protocols for achieving different functions. A protocol stack for the multi-UAV system is presented in Fig. \ref{fig:protocol} to specify potential protocols considering the proposed application. It shows a five layers structure, application layer, transport layer, network layer, data link layer and physical layer. For \textcolor{black}{a} UAV system, most of the communication is handled by the robotic middleware layer, which is part of the application layer. It provides APIs for discovery, publish/subscribe mechanisms and to configure the communication \ac{QoS} and the multicast service. Considering the compatibility to robotic middleware, TCP and UDP are addressed in the transport layer.  In addition, IP and Multicast function are addressed in the \textcolor{black}{network} layer. \textcolor{black}{A multicast} function is involved to support node discovery mechanism in the network as well as general multicast data transmission. To maintain reliable and flexible network connectivity, \textcolor{black}{a} mesh network protocol is considered in the \textcolor{black}{network} layer. In \textcolor{black}{the} data link and physical layer, different wireless technologies, such as WiFi, LTE, 5G and LoRa, are in the investigation scope. }

\section{Wireless Technologies for UAV Networks \label{sec:wireless}} 
\subsection{{Wireless Standards}}
{Various wireless standards exist or show potential in \textcolor{black}{the} UAV industry. Wireless standards come as a reasonable approach since the communicating nodes exchange and/or report data about possibly location, velocity, energy consumption, among other variables. In what follows, we describe different wireless technologies that are developing for supporting cooperative \ac{UAV} applications.}



\subsubsection{3GPP Cellular Networks}
The \ac{3GPP} standards have been the market-based evolution of cellular technical requirements, technologies and services.  Cellular networks typically make use of licensed spectrum, {culminating in Release 10 utilizing sub-6 GHz bands \cite{lte_3gpp_band} with \ac{LTE-A}. Later, Release 12 from the beginning in 2012 was marked by the \ac{ProSe} specifications \cite{prose_specs1,prose_specs2} for \ac{D2D} communications, such as LTE Direct, for \ac{UE} discovery, relaying, and signaling without explicit network access.}
Due to the more recent expansion of the \ac{IoT}, 3GPP has developed in Release 13 low power wide area radio standards leading to technologies such as LTE-M,  addressing \ac{eMTC} and \ac{NB-IoT} \cite{realease13_3gpp}. 
{Since} Release 15 {from} 2018, 3GPP enhancements \textcolor{black}{included} 5G support for aerial vehicles\textcolor{black}{. It launched} a dedicated study to investigate the ability for aerial vehicles to be served using LTE network deployments with base station antennas targeting terrestrial coverage. {5G NR includes new frequency bands known as mmWave bands for \ac{UAV} applications. Examples of these frequencies are e.g., 26 GHz, 28 GHz, 39 GHz (licensed) and 60 GHz (unlicensed, based in WiFi) \cite{5g_uav_survey}, which are in the Ka and V bands of the electromagnetic spectrum.}
{In an measurement report from Qualcomm, a UAV can be controlled in \ac{BVLOS} at the distance of 6 km to the serving cellular base station and up to 120 m above the ground} \cite{QualcommTechnologies2017}. 
More recently, 5G specifications show high potential for {supporting} UAV applications {with the advantages of} lower latency and higher data rate \cite{WirenEricsson2017}.

\subsubsection{Wireless Local Area Networks}
The IEEE 802.11 standard series have proved as an important de facto family of standards in \acp{WLAN} given their coverage, ease of deployment, achievable rates, and maintenance capabilities. 
{The standards provide a \textcolor{black}{basis} for \ac{WiFi} systems.}
Most WiFi variants rely on the unlicensed spectrum {that is posed on} 2.4 GHz (b/g) and 5.8 GHz (a), where they provide a range of up to hundreds of meters and throughput in tens of Mbps \cite{Yanmaz2013, FromGroundAsadpour2013}. {Considering the usage in UAV systems}, several variants have been identified in the scope of \textcolor{black}{the} survey \cite{Zolanvari2020}. The IEEE 802.11a applies \ac{OFDM} with high data rates and signal resistance to {address} multipath {effects}, but shorter distance {while comparing with} the IEEE 802.11b {that} uses \ac{DSSS}. The IEEE 802.11g variant was designed to inherit the advantage of \textcolor{black}{the} previous two standards on rate and range, whereas the IEEE 802.11n is improved by using \ac{MIMO} to exploit the diversity given by transmitting and receiving antennas at the same time.
{Improved \textcolor{black}{from} \ac{WAVE}, the IEEE 802.11p is designed targeting \ac{V2X} communications to provide better support in different intelligent transport services and applications. To utilize the benefits of mesh networking, the IEEE 802.11s is proposed. Additionally, the IEEE 802.11ah is designed to \textcolor{black}{respond} the requirements from IoT use cases by extending the range with the support of sub-GHz bands and  \ac{MIMO}.
}

{\subsubsection{Wireless Personal Area Networks}}
{Communication products following the IEEE 802.15.4 standard are \textcolor{black}{relatively common in the UAV} accessory market. The IEEE 802.15.4 standard aims to offer low-power low-cost personal area networks with \ac{D2D} communication. However, there is often limited data rate compared with products under the IEEE 802.11 series. }
On the other hand, the IEEE 802.15.4 with Ultra-Wide Band (UWB known as IEEE 802.15.4a) supports \textcolor{black}{a} higher data rate than the 802.15.4 standard from 2006.
For instance, DecaWave's  UWB transceiver claims the maximum data rate up to 6.8 Mbps and the extended communication range up to 290 m, at 110 kbps, 10\% \ac{PER} \cite{Jimenez2016}. Schmidt et al. \cite{Schmidt2018} presented an experimental study of UWB \textcolor{black}{compared} with ZigBee, which is \textcolor{black}{based} on 802.15.4, in an indoor multipath propagation environment. It showed that UWB outperforms ZigBee in terms of range, data rate and \textcolor{black}{packet} loss rate. For UAV A2G propagation channels, Khawaja et al. \cite{Khawaja2016, Chen2018} investigated \textcolor{black}{the} UWB channel model for path loss, multipath and characterization of A2G channels on various UAV distances and heights.
{UWB devices are widely adopted to provide indoor positioning services but limited literature can be found for \textcolor{black}{UAV-related} communications. Although the technology shows interesting \textcolor{black}{possibilities} on industrial communication, researchers should also be aware \textcolor{black}{of} the extra constraints from the current regulations to UWB devices. In normal cases, UWB devices are required to be used with very low radiated power to control its interference to other devices, whose frequency bands are overlapped by the UWB device. For using UWB in Europe, regulations can be found in ETSI EN 302 065-1 \cite{European2016},} which states the maximum mean power spectral density should not exceed -41.3 dBm/MHz at the preferred range of operating bandwidth from 3.1 GHz to 4.8 GHz and from 6 GHz to 9 GHz. Besides, more sector-specific regulations will be taken into consideration when the UWB transmitter equipment is installed at a fixed outdoor location for use in flying models and other forms of aviation \cite{European2016}.

{\subsubsection{Low Power Wide Area Networks}}
{Targeting IoT applications, Cycleo acquired by Semtech \cite{semtech} developed a set of standards named \ac{LoRa} and \ac{LoRaWAN}.} LoRa {defines} the \ac{PHY} aspects of the protocol stack, whereas LoRaWAN deals with the data link and upper protocol layers. LoRa {can be utilized} in both licensed (169 MHz, 868 MHz) and unlicensed spectrum (433 MHz, 914 MHz and 2.4 GHz), which makes it {a proper candidate} for planned as well as ad-hoc deployments. LoRa {is supported by} \ac{CSS} modulation, another spectral expansion technique adapted from \ac{DSSS}, where a narrowband signal with a given data rate is chipped at a higher rate, known as the chip rate, and then modulated into a frequency-varying signal known as a chirp \cite{loraAugustin2016}. LoRa diminishes the throughput for longer ranges. 
LoRaWAN {can offer} a raw maximum throughput of 27 kbps. \ac{LoRaWAN} operates \textcolor{black}{with} a random access protocol such as ALOHA in a star-of-stars topology with three device classes that trade connection event-rate against battery lifetime \cite{Adelantado2017}. {There are three categories of LoRa devices, named A, B and C. Class A offers the lowest energy consumption but supports only low downlink requirements. It is a baseline implementation for all devices that deployed LoRa. Second, Class B is designed to support extra downlink requirements and traffic scheduling by sending periodic beacons from a local gateway. Class B contains all features in Class A. Third, Class C \textcolor{black}{is} continuously} listening to \textcolor{black}{the} wireless medium except when sending data. Class B and C cannot be implemented in the same device, but all three classes can exist simultaneously in a given network. {Note, the duty-cycle of IoT devices is regulated}, which defines the maximum amount of time that a device can occupy a transmission channel\textcolor{black}{. It} is set to 1\% in the 868 MHz band across the EU \cite{ETSI}.  

\begin{table}[]
\centering
\caption{Wireless technologies for UAV applications: In-field measurements.}
\label{table:In_field_measurements}
\begin{threeparttable}

\begin{tabular}{|>{\centering}p{0.14\textwidth}|>{\centering}p{0.17\textwidth}|>{\centering}p{0.12\textwidth}|>{\centering}p{0.18\textwidth}|>{\centering}p{0.09\textwidth}|>{\centering}p{0.08\textwidth}|}
\hline 
\textbf{Network Type} & \textbf{Technology} & \textbf{Link Type} & \textbf{Throughput (Mbps)} & \textbf{Range (m)} & \textbf{Refs.}\tabularnewline
\hline 
\multirow{4}{0.09\textwidth}{Cellular} & 3G & A2G, G2A & 0.384 & 100 & \cite{Zhu2015}\tabularnewline
\cline{2-6} \cline{3-6} \cline{4-6} \cline{5-6} \cline{6-6} 
 & LTE & A2G, G2A & 0.1-18.5 & 20-4500 & \cite{LTEAmorim2017} \cite{LTEKovacs2018} \cite{LTE2Kovacs2018} \cite{QualcommTechnologies2017} \tabularnewline
\cline{2-6} \cline{3-6} \cline{4-6} \cline{5-6} \cline{6-6} 
 & LTE-M & Drive test & 0.1 & 15000 & \cite{Lauridsen2016}\tabularnewline
\cline{2-6} \cline{3-6} \cline{4-6} \cline{5-6} \cline{6-6} 
 & NB-IoT & Drive test & 0.2 & 15000 & \cite{Lauridsen2016}\tabularnewline
\hline
\multirow{6}{0.09\textwidth}{WLAN} 
 & 802.11b & A2G & 1.2 & 2000 & \cite{Brown}\tabularnewline
\cline{2-6} \cline{3-6} \cline{4-6} \cline{5-6} \cline{6-6} 
 & 802.11a & A2G, G2A, A2A & 8-14 & 300-{500} & \cite{Yanmaz2013} \cite{ExperimentalYanmaz2014}
  {\cite{6953010}}\tabularnewline
\cline{2-6} \cline{3-6} \cline{4-6} \cline{5-6} \cline{6-6} 
 & 802.11n & A2A & 2.2-20 & 80-300 & \cite{Asadpour2013} \cite{FromGroundAsadpour2013}\tabularnewline
\cline{2-6} \cline{3-6} \cline{4-6} \cline{5-6} \cline{6-6} 
 & 802.11g & G2A, A2G, A2A & 8 & 75 & \cite{Morgenthaler2012}\tabularnewline
\cline{2-6} \cline{3-6} \cline{4-6} \cline{5-6} \cline{6-6} 
 & 802.11p & V2V & 3 & 400 & \cite{Lv2016}\tabularnewline
\cline{2-6} \cline{3-6} \cline{4-6} \cline{5-6} \cline{6-6} 
 & 802.11ah & Outdoor moving & 0.15 & 300 & \cite{outdoorBellekens2017}\tabularnewline
\hline 
\multirow{3}{0.09\textwidth}{WPAN} & 802.15.4 & A2G, A2A & 0.056-0.250 & 500-1500 & \cite{Allred2007}\cite{Asadpour2013}\tabularnewline
\hline
\multirow{2}{0.09\textwidth}{LPWAN} & LoRa 868 MHz, 915 MHz & Drive test & 0.006 & 650-1600 & \cite{loraAugustin2016}\cite{Rahmadhani2018}\tabularnewline
 \cline{2-6} \cline{3-6} \cline{4-6} \cline{5-6} \cline{6-6} 
\hline
\end{tabular}
\begin{tablenotes}
	\item V2V: Vehicle to Vehicle (on \textcolor{black}{the} ground) communication
	\item Drive test: The testbed is placed on a car to test the communication performance when driving on the road \textcolor{black}{at different speeds}.
\end{tablenotes}
\end{threeparttable}
\end{table}

\subsection{In-Field Performance Measurements}
{Different performance measurements in regards to existing wireless technologies with the use case related to UAVs are reviewed and compared.
We focus on reviewing the in-field measurement results of throughput and communication range, which are critical metrics while considering the use case of the UAV~\cite{Andre2014}. In addition, the type of the communication link is listed, which specifies involved transmitters and receivers. As shown in Table~\ref{table:In_field_measurements}, UAV experimental data from different wireless technologies are summarized in items of range and throughput.
}

{According to the data in} Table \ref{table:In_field_measurements}, {it can be observed that WiFi has been tested relative often }for the purpose of UAV communications since 2007 \cite{Brown, outdoorBellekens2017}. With the appearance of IEEE 802.11s supporting meshed topologies, evaluations for \ac{A2A} multi-UAV communication occurred as well \cite{Morgenthaler2012}, \cite{ExperimentalYanmaz2014}.  
{As an amendment for vehicular wireless usage~\cite{Lv2016}, experiments related with IEEE 802.11p are reviewed}
In \cite{outdoorBellekens2017}, Bellekens et al. present their measurements on IEEE 802.11ah in an outdoor moving scenario to \textcolor{black}{evaluate} its improvement on the range. 
{In the category of IEEE 802.15.4, XBee Pro \cite{xbee} has been adopted }in many projects and consumer UAVs for the communication between the UAV and the ground station \cite{Asadpour2013}, \cite{Allred2007}. { XBee developed by Digi is a branch of the ZigBee with enhanced transmission power. It targets the application that requires} low latency and predictable communication timing.

In recent years, {solutions \textcolor{black}{that use} cellular services, such as LTE, are considered and tested to be deployed in the UAV system~\cite{LTEAmorim2017}. The \textcolor{black}{cellular-based} technology offers the capability to intensively extend the range of communication for UAVs to be controlled} in a \ac{BVLOS} fashion {under the support of cellular infrastructures} \cite{QualcommTechnologies2017}. {As can be seen, LTE provides sufficient throughput for the use cases that \textcolor{black}{require}} live image and video transmission \cite{LTE2Kovacs2018}.
{Lately, LoRa is taken as a new option for UAV communication.} Rahmadhani et al.~\cite{Rahmadhani2018} {evaluated} LoRa as a {redundant} telemetry system for \textcolor{black}{UAVs}. A LoRa based communication system for swarm control is also implemented \cite{Yuan2018}. 


{Apart from throughput and range, \textcolor{black}{the }communication delay \textcolor{black}{is also} evaluated within each wireless technology.} Yanmaz et al.~\cite{Yanmaz2013} reported that the packet inter-arrival time of IEEE 802.11a is between 0.1 ms to 10 ms and 90\% of\textcolor{black}{ packets are }less than 1 ms. Packet interarrival time of IEEE 802.11p {are tested in \cite{Lv2016}, which \textcolor{black}{claims}} 80\% of the packets are \textcolor{black}{within} 100 ms and 99\% in 400 ms. {Network delay of 3G technology \textcolor{black}{is} measured in \cite{Zhu2015}, where reports} 38\% of the packets {can be delivered} in 100 ms {and 95\% are within} 200 ms.  In the case of LTE, Kovacs et al. \cite{LTE2Kovacs2018} {reports} that approximately 75\% of \textcolor{black}{the} delay per packet is less than 30 ms. However, { the delay shows an oblivious increase} when the \ac{RSRQ} (a type of channel-to-interference ratio) {drops} lower than -20 dB. Qualcomm \cite{QualcommTechnologies2017} measured the delay in \textcolor{black}{handovers} {between LTE services}. 80\% of \textcolor{black}{the} delays are less than 40ms. {Furthermore,} Lauridsen et al. \cite{Lauridsen2016} {present} the average session delay of LTE-M (200 ms) and NB-IoT (1000 ms) {that are tested in the rural area}.

{Besides the performance measurements, UAV communication \textcolor{black}{channels} are also researched. Naturally, there are two types of communication \textcolor{black}{channels: the air-to-air channel and air-to-ground channel}.}
We {surveyed} UAV related channel modeling studies that use practical measurements for the model optimization {and validation}. {In} IEEE 802.11 series, Goddemeier et al. \cite{Goddemeier2015} extended the Rice channel model to account for multipath effects introduced by UAVs for an A2A channel. Yanmaz et al. \cite{Yanmaz2011} {present} A2G channel {modeling with validation} measurements. Kung et al. \cite{Kung2010} measured the diversity of a UAV in a G2A mesh network. For IEEE 802.15.4, Ahmed et al. \cite{Ahmed2016} characterized A2A, A2G and G2A channels in aerial wireless sensor networks. For LTE, Amorim et al. \cite{LTEAmorim2017} {proposed} models for path loss exponents and shadowing between UAVs and cellular networks. Al-Hourani et al. \cite{Al-Hourani2018} {obtained a model to represent} the statistical behavior of the path loss from a cellular base station to a UAV. {Finally,} Afonso et al. \cite{Afonso2016} {investigated} A2G channel with {several} performance measurements obtained {through} the flight tests {using different} cellular technologies, EDGE, HSPA+ and LTE.

{Channel modeling for UAV communication are mainly considering WiFi technologies and Cellular technologies.} It is observed that for A2A channel, IEEE 802.11 WiFi technology has been researched the most. For the A2G/G2A channel, research papers for LTE, IEEE 802.11 and IEEE 802.15.4 have presented insightful contributions on their channel modeling.
Although LoRa shows the possibility in the UAV application, we have found very few works regarding its channel modeling.

{One emerging alternative for WLAN that has appeared is a LoRa variant on 2.4 GHz \textcolor{black}{was} initially commercialized by Semtech as well. This technology operates under the same principles as the original specification but avoids the rate and duty cycle limitations of sub-GHz counterparts. Reported theoretical rates are in the order of 250 kbps \cite{semtech}. Despite these benefits, more evaluations are needed to understand the impact of the interference coming from the \textcolor{black}{bands used} in IEEE 802.11. At \textcolor{black}{the} moment of this review, no reported measurements on LoRa at 2.4 GHz have been identified in the literature so far.}


%
%
\section{Wireless Mesh Networks for UAVs
\label{sec:meshnetworks}}
{
A mesh network is a communication network topology in which infrastructure nodes connect directly, dynamically and often non-hierarchically to as many other nodes as possible for multi-hop packet forwarding.
Various interconnected nodes allow \textcolor{black}{establishing} paths to efficiently route data across them and messages pass through intermediate nodes from any given source to a specific destination in multiple hops.
The benefits of mesh networks \textcolor{black}{includes a rather} rapid installation and low maintenance costs. 
Also, mesh networks can add robustness and eliminate a single point of failure due to their decentralized network architecture.
However, wireless mesh networks are prone to link failures due to interference, mobility and may fall short to meet data rate demands.
}

{
A few studies of wireless mesh networks for \acp{UAV} have been reported in the  literature~\cite{6477825,6953010,6125375}.
In~\cite{6477825}, the authors evaluated a framework for an adaptive and mobile wireless mesh network using small \acp{UAV}.
The network was based on IEEE~802.11s to provide a WiFi mesh network.
The IEEE~802.11s amendment has been adopted in the IEEE~802.11 standard since 2012~\cite{802.11-2012}. 
It brings important methods for bridging the path selection to the \ac{MAC} layer and making \ac{PHY} data easily available for routing optimization.
The work demonstrated how each of the mesh nodes acted as a wireless access point and offered access to  802.11g network services.
This allowed \textcolor{black}{an} extended communication range between end-points through the mesh infrastructure, by using \acp{UAV} as air relay nodes in the mesh~\cite{6477825}.
In~\cite{6953010}, the authors provided an experimental study of two-hop communication using 802.11s mesh system. The study reports on network connectivity ranges up to 500 m with 12~Mbps. However, they also observed that the ``mesh  extension  802.11s  is  only  moderately  suited  for  networking  UAVs'' as the system only switched to two-hops for 3\% of the transmitted packets \cite{6953010}.   
This urges further research in 802.11 mesh networks for UAVs. 
}

{Mesh networks dynamically self-organize and self-configure. 
To achieve this operation, mesh networks run discovery and peering protocols to locate other nodes and to manage membership of the mesh network.
This is handled by a \ac{MPM} mechanism. 
Essentially, a joining node has to discover the operative wireless mesh network by scanning all radio channels and waiting for beacons (passive mode) or by issuing beacon requests and awaiting beacon responses (active mode). 
Direct communication between neighbor nodes is allowed only when they are peer mesh nodes. 
After a mesh discovery, two neighbor mesh nodes may agree to establish a mesh peering to each other.
}

{
The IEEE~802.11s amendment adds an \ac{MPM} protocol \textcolor{black}{that} facilitates the establishment and closure of the mesh peering~\cite{802.11-2012}. 
Generally speaking, \ac{MPM} \emph{request} messages are used by a \ac{SME} function to establish, confirm, or close a mesh peering with other peering nodes.
These peerings are managed by the mesh nodes through the \ac{MLME} function.
The \ac{MPM} \emph{confirm} message reports the results of a request.
The \ac{MPM} \emph{indication} message is used by the \ac{MLME} to report any peering states with other nodes to the \ac{SME}. 
Finally, \ac{MPM} \emph{response} messages are used to send responses to the \ac{MLME} specified by the peer node \ac{MAC} addresses.
}

{Mesh networks relay messages using either a flooding technique or a routing mechanism.
Messages are forwarded along a path from a sender to a receiver node.
Nodes are making forwarding decisions based on path information of the network. 
Basic forwarding information consists of the destination mesh node \ac{MAC} address, a next-hop address, a precursor list, and the lifetime of forwarding information. 
A mesh path selection protocol may benefit from combining reactive and proactive elements that \textcolor{black}{enable} efficient path selection in a wide variety of mesh networks.
Although agnostic to any specific routing protocol, the IEEE~802.11s mesh network standard promotes the \ac{HWMP} as the preferred routing protocol~\cite{802.11-2012}.
\ac{HWMP} is a hybrid protocol that combines the flexibility of an on-demand path selection process with proactive topology tree extensions.
The reactive part is inspired by the \ac{AODV} protocol~\cite{rfc3561} adapted for MAC address-based path selection and link metric awareness. 
Two modes of operation exist: 1) the on-demand mode allows mesh nodes to communicate using peer-to-peer paths and 2) the proactive tree building mode provides a tree building functionality added to the on-demand mode. 
Path information is maintained by use of four protocols messages: PREQ (path request), PREP (path reply), PERR (path error) and RANN (root  announcement).
}
{If a source mesh node needs to find a path to a destination mesh node.
It broadcasts a PREQ with the path target specified in a list of targets.
When a mesh node receives a new PREQ, it creates or updates its path information to the originator mesh
node and propagates the PREQ to its neighbor peer mesh nodes.
After creating or updating a path to the originator mesh node, the target mesh node sends an individually addressed PREP back to the originator mesh node.
If the mesh node that received a PREQ is the target mesh node, it sends an individually addressed PREP back to the originator mesh node after creating or updating a path to the originator.
In the proactive mode, the root mesh node periodically propagates RANN messages in the network. 
The RANN messages are used to disseminate path metrics to the root mesh node and do not establish path information.
Upon reception of a RANN, each mesh node that has created or refreshed a path to the root mesh node sends an individually addressed PREQ to the root mesh node via the mesh node from which it received the RANN.
}

{A few studies of the use of \textcolor{black}{dynamic} routing protocols for \ac{UAV} mesh networks have been reported \cite{6125375, 7980778}. 
In \cite{6125375}, the authors studied the performance of four available mesh routing protocol implementations (open80211s, BATMAN, BATMAN Advanced and OLSR) in the context of swarming applications for UAVs. The paper evaluated the performance of IEEE 802.11s with emphasis on goodput and the transmission delay.
In \cite{7980778} the authors simulated the performance of \ac{HWMP} and suggested an optimization that implemented the proactive tree-based routing scheme applied on a ground infrastructure, while the reactive routing is initiated by the mobile mesh node.
The study showed a significant reduction in routing overhead alongside with improvements of transmission delay and \ac{PSR}.
Other studies confirm that pure reactive routing scheme may lead to a significant degradation of network performance in terms of added delay and reduced goodput when attempted to be adapted to highly mobile network scenarios such as satellite constellation networks~\cite{hernandezmarcano2020}. These protocols are summarized in Table~\ref{table:meshtable} below.
}

\begin{table}[h!]
\footnotesize
\caption{Surveyed Networking Protocols}

\begin{tabular}{|>{\raggedright}p{0.18\columnwidth}|>{\raggedright}p{0.18\columnwidth}|>{\raggedright}p{0.16\columnwidth}|>{\raggedright}p{0.20\columnwidth}|>{\raggedright}p{0.10\columnwidth}|}
\hline 
\textbf{Protocol} & \textbf{Type} & \textbf{Reactiveness} & \textbf{Metrics} & \textbf{Refs.} \tabularnewline
\hline 
HWMP (802.11s) & Hybrid & Hybrid & Range, PSR, Delay, Overhead & \cite{6477825,6953010,6125375,7980778} \tabularnewline
\hline 
BATMAN(-adv) & Link State & Proactive & Goodput, Delay & \cite{6125375} \tabularnewline
\hline 
OLSR & Link State & Proactive & Goodput, Delay &  \cite{6125375} \tabularnewline
\hline 
AODV & Distance Vector & Reactive & PSR, Goodput & \cite{hernandezmarcano2020} \tabularnewline
\hline 
\end{tabular}
\label{table:meshtable}
\end{table}

\vspace{10mm}



%
\section{Communication Middleware \label{sec:rosnetworks}}


{
As can be observed, the majority of commercial and industrial autonomous UAVs are developed based on \ac{ROS} or its branches. Therefore communication protocols and interfaces that adopted in this communication middleware are investigated to identify pros and cons of the methods that address fundamental functions including connectivity management and data distribution. }
\subsection{API and Protocol}
\begin{figure}[t]
 \centering
 \includegraphics[width=\columnwidth]{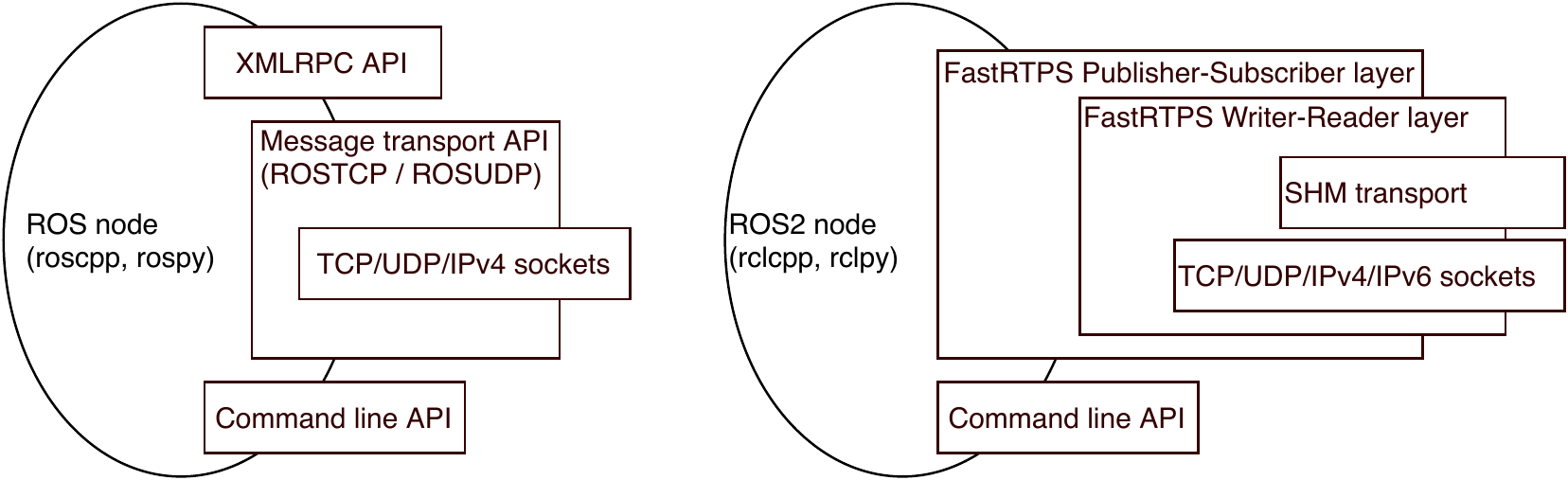}
 \caption{ROS node APIs and ROS2 node APIs.}
 \label{fig:ros_node_api}
\end{figure}

{In ROS based systems, standard APIs and protocols are provided, which allows \textcolor{black}{ wide reuse of robotic software components}. In this system, communication usually starts \textcolor{black}{from} one ROS node to another or a few other ROS nodes, which can be in the local UAV or in other UAVs (remote nodes). ROS nodes (e.g. roscpp, rospy) are implementations of software or processes that are doing on-board (UAV) processing. ROS nodes have to be developed by following its standard APIs that \textcolor{black}{define} the data format (i.e. ROS messages) as well as related transport protocols. 
\ac{ROS} uses ROS messages for communication between ROS nodes based on a publish and subscribe mechanism. The mechanism is implemented via XMLRPC for naming and registration services, which is a remote procedure call (RPC) protocol that uses XML to encode its calls and HTTP as a transport mechanism. 
However, the use of RPC introduces a risk of limited availability due to its blocking nature in contrast to the more loosely coupled publish-subscribe model that does not require simultaneous availability of subsystems.
}
{
ROS nodes can be registered as publishers, subscribers, and service providers. The mechanism uses ROSTCP and ROSUDP for transportation, which is based on standard TCP/IP or UDP/IP sockets. There are three standard APIs for a ROS node, including XMLRPC API, ROS message transport protocol implementation (ROSTCP and ROSUDP) and command line API. Fig.~\ref{fig:ros_node_api} illustrates a ROS node and its three standard public interfaces, of which the message transport API is implemented based on standard socket supporting TCP, UDP and IPv4 addresses.}

\begin{figure}[t!]
 \centering
 \includegraphics[scale=0.7]{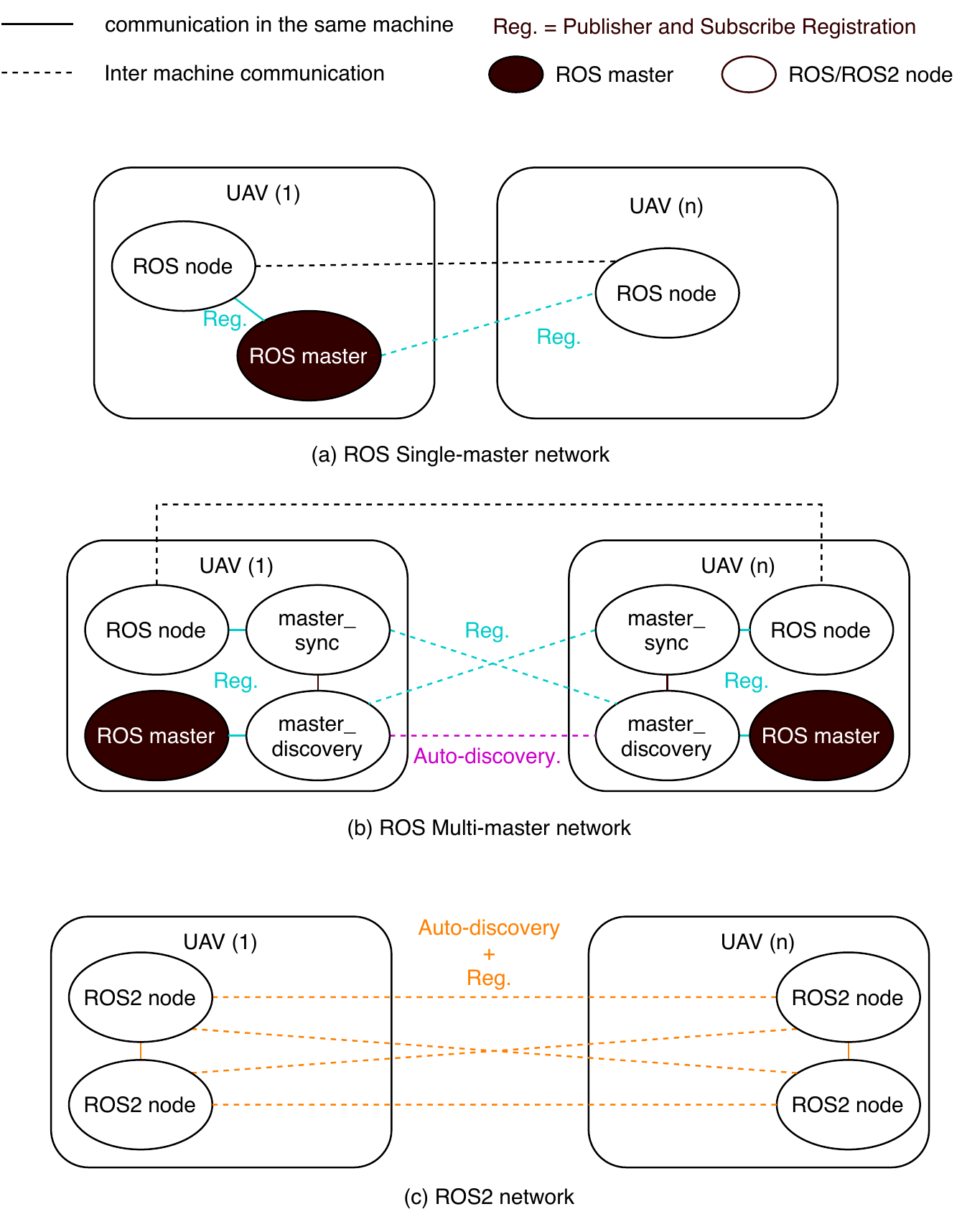}
 \caption{Three types of ROS network: (a) ROS Single Master, (b) ROS Multi-master, (c) ROS2.  }
 \label{fig:ros_network}
\end{figure}
\subsection{Network}
{
As shown in Fig. \ref{fig:ros_network}, there are three types of \textcolor{black}{networks related} with the ROS, i.e., single master ROS network~\cite{Quigley2009}, multi-master ROS network~\cite{Tiderko2016} and ROS2 network~\cite{Maruyama2016}. In some applications, different ROS networks are deployed in combination~\cite{Eros2019}. A ROS master provides naming and registration services to the rest of the nodes in the ROS. The role of the master is to enable individual ROS nodes to locate one another. Once they are located by each other, they will communicate peer-to-peer.
In \textcolor{black}{a} single master ROS network, only one ROS master is running on a host. ROS nodes for different functions are running on the \textcolor{black}{local-host} and remote hosts. Wireless or wired communication \textcolor{black}{is} required for communication to remote ROS nodes. A single master ROS network uses centralized connectivity management. Therefore the complete line of communication is prone to failures if the master fails.
}
{
Multi-master ROS network provides an extension for \textcolor{black}{a} single master ROS network to spread the load of the centralized ROS master. It involves multiple ROS masters for connectivity management in the network, e.g. each host has its own ROS master, Fig.~\ref{fig:ros_network}. Tiderko et al.~\cite{Tiderko2016} published a ROS package named $\texttt{multimaster\_fkie}$ to help quickly setup a multi-master ROS network for a team of robots. For discovering remote hosts (their ROS masters) in the network, $\texttt{multimaster\_fkie}$ provides $\texttt{master\_discovery}$ ROS node which supports either multicast \ac{UDP} or \ac{zeroconf} protocol. In \textcolor{black}{the} multicast \ac{UDP} \textcolor{black}{mode}, heartbeat messages are sent by each remote hosts to notify the update of its state as well as measuring link quality of the communication channel. For ROS message distribution and integration, $\texttt{multimaster\_fkie}$ provides $\texttt{master\_sync}$ node to register and unregister topics and services on behalf of the respective nodes. $\texttt{master\_sync}$ queries the current state from the local $\texttt{master\_discovery}$ node and then updates its local ROS master. Therefore the registration can be managed locally instead of using the remote ROS master. After the registration, data transportation between ROS nodes uses peer-to-peer communication based on TCP/IP or UDP/IP.
}

\begin{table}[b]
\small
\caption{Concept mapping between ROS2 and DDS middleware}
\begin{tabular}{|c|c|c|}
\hline 
\multirow{2}{*}{\textbf{ROS2}} & \multicolumn{2}{c|}{\textbf{DDS middleware (FastRTPS)}}\tabularnewline
\cline{2-3} \cline{3-3} 
 & \textbf{Publisher-Subscriber layer} & \textbf{Writer-Reader layer}\tabularnewline
\hline 
Node & Participant & RTPSParticipant\tabularnewline
\hline 
Publisher & Publisher & RTPSWriter\tabularnewline
\hline 
Subscriber & Subscriber & RTPSReader\tabularnewline
\hline 
Service & Publisher + Subscriber & RTPSWriter + RTPSReader\tabularnewline
\hline 
Client & Publisher + Subscriber & RTPSWriter + RTPSReader\tabularnewline
\hline 
\end{tabular}
\label{table:concept-mapping}
\end{table}

{A ROS2 network is different from a ROS network. The ROS master no longer exists in a ROS2 network. A ROS2 network is implemented by using \ac{DDS}/\ac{RTPS} as its middleware, which provides discovery, message definition, message serialization and publish-subscribe transportation. ROS2 provides a ROS type of interface on top of DDS, which hides most of the complexity of DDS for ROS users. Access to DDS specific API is provided separately in the case for the need to integrate with existing DDS system, e.g., eProsima FastRTPS, OpenSpliceDDS and CycloneDDS \cite{Maruyama2016}. FastRTPS is the default middleware implementation in ROS2. The concept mapping between ROS2 entities and FastRTPS entity is shown as Table \ref{table:concept-mapping}. 
FastRTPS implements five transports, i.e., UDPv4, UDPv6, TCPv4, TCPv6 and shared memory (SHM) cf. Fig.~\ref{fig:ros_node_api}. 
By default, SHM will be used for communication between ROS2 nodes in the same machine. UDPv4 is used by default for inter-machine communications. 
Note the TCP and UDP transport concepts (server and the client) are independent \textcolor{black}{of} the FastRTPS concepts (Publisher, Subscriber, Writer, and Reader). Any of FastRTPS concepts can operate as a TCP/UDP Server or Client as these entities are used only to establish the TCP/UDP connection and the FastRTPS protocol works over it. FastRTPS uses a configurable \textcolor{black}{two-phase} discovery, Participant Discovery Phase (PDP) and Endpoint Discovery Phase (EDP). In PDP, each participant sends periodic announcement messages using multicast addresses or following a list of configured target addresses. In EDP, participants share information of their publishers and subscribers with each other by using channels established during PDP. 
A ROS2 network provides more configuration options to maintain the communication \ac{QoS}, including communication data flow control, message source priority setting, message deadline setting and message filter. Durability configuration with history memory policy is also provided for handling messages that exit on the topic before a subscriber joins. 
}
{ROS2 provides a package, $\texttt{ros1\_bridge}$, to support bidirectional communication between ROS and ROS2 because the majority of robotic functions (e.g. control, localization and mapping) have been implemented and fully validated as ROS nodes using standard ROS APIs. \textcolor{black}{To} control the risk of cross-platform deployment, ROS2 is often built as a parallel set of packages that may be installed alongside and interoperate with ROS.
}

\begin{table}[h!]
\footnotesize
\caption{Comparison of Advantages and Disadvantages in different ROS networks}

\begin{tabular}{|>{\raggedright}p{0.15\columnwidth}|>{\raggedright}p{0.39\columnwidth}|>{\raggedright}p{0.39\columnwidth}|}
\hline 
\textbf{Network} & \textbf{Advantages} & \textbf{Disadvantages}\tabularnewline
\hline 
Single-master ROS network & - Simple system structure

- Seamless integration with \textcolor{black}{a} large amount of third-party software (ROS
package) for different robotic functions & - Depends on a centralized device

- Very limited network scalability

- Limited extendability of ROS communication middleware with closely
coupled structure

- Cumbersome network configuration by default

- Not support discovery in network by default

- Limited options to control communication QoS\tabularnewline
\hline 
Multi-master ROS network & - Relax the dependence on a centralized device

- Seamless integration with \textcolor{black}{a} large amount of third-party software (ROS
package) for different robotic functions

- Improved network configuration process with the support of third-party
ROS packages

- Support discovery in \textcolor{black}{the} network by third-party ROS packages & - Limited network scalability

- Limited extendability of ROS communication middleware with closely
coupled structure

- Limited options to control communication QoS\tabularnewline
\hline 
ROS2 network & - Flexible network topology

- Good network scalability

- Good extendability of ROS2 communication middleware with hierarchical
architecture

- Simple network configuration

- Support discovery in network

- Support reliable communication on top of UDP

- More options to control communication QoS & - Lack of third-party ROS2 packages

- Extra time cost for message conversion to fit DDS requirements\tabularnewline
\hline 
\end{tabular}
\label{table:proscons}
\end{table}

\subsection{Performance}
{\textcolor{black}{The pros and cons concerning} communication latency and reliability, network scalability and software extendability in three different ROS networks are summarized in Table~\ref{table:proscons}.}
{Many projects developed to evaluate and improve the performance of \ac{ROS} and \ac{ROS2}. 
Sticha in \cite{Sticha2014} measured the real-time capabilities of the ROS communication middleware. To provide reliable and short latency, a real-time ROS architecture, RT-ROS~\cite{Wei2016}, is proposed with \textcolor{black}{a real-time node and a non-real-time node} that can be respectively run on a real-time OS as well as Linux. Measurements of the latency and the diversity of the latency using ROS communication middleware \textcolor{black}{are} presented by \cite{Saito2016}. \textcolor{black}{ Accordingly, a priority-based} message transmission mechanism with synchronization system is proposed~\cite{Saito2019}. Tardioli et al.~\cite{Tardioli2019} evaluated the ROS communication middleware with wireless communication in presence of multiple \textcolor{black}{message-flows}, different message sizes and frequencies. An optimized ROS node (Pound) was proposed to reduce the latency and jitter in wireless multi-master ROS network.}

{In 2020, Jiang et al.~\cite{Jiang2020} evaluated the interprocess communication latency caused by messaged serialization (de-serialization), convert (de-convert) and transport in a ROS2 system. Convert is a process to transform messages used by the application (e.g. programming language interface layer of Python or C++) to satisfy DDS requirements. { It is observed that default convert and de-convert in ROS2 middleware layer are accounted for the majority \textcolor{black}{of the} time cost (e.g. 89.1\%), especially in the complex message (e.g multi-dimensional array with different data types).  \textcolor{black}{The time} cost of convert and de-convert is linearly related to the complexity of the message data structure. To control the complexity of conversion, serialization is considered to be handled in advance in the programming language interface layer. An adaptive two-layer serialization algorithm was then proposed.}
Maruyama et al.~\cite{Maruyama2016} tested the performance of ROS and ROS2 in 2016 on different aspects including end-to-end latency, throughput, number of threads and memory consumption. On both ROS and ROS2, they evaluated intra-process communication and inter-process communication in a single machine and two machines with an Ethernet connection. The data link for connecting ROS and ROS2 using $\texttt{ros1\_bridge}$ was evaluated. In addition, the influence of switching DDS implementations (Connext, OpenSplice and FastRTPS) and the \ac{QoS} Policies in ROS2 are measured. In 2019, for the purpose to easily compare the performance, FastRTPS automated benchmark \cite{eprosima} is published with gathered performance results for different DDS implementations and versions. Another project, $\texttt{ros2\_performance}$ \cite{irobot}, is published for ROS2 performance evaluation with customisable nodes configuration (e.g. number of nodes and their publish/subscribe relations) and automated measurements report. 
}

\section{Conclusions \label{sec:conclusion}}
{This paper presents an application analysis and a survey of communication solutions for multi-UAV system design, in the aspect of existing wireless technologies, the mesh network management and the robotic communication middleware. Features and dependencies of each candidate solution in these three aspects are identified following the proposed protocol stack, Fig.~\ref{fig:protocol}.
The dependency due to APIs or function implementations limits the choice of solutions for the target communication system, i.e., \textcolor{black}{when one solution is selected}, multiple other dependent solutions are decided as well, even though they are not the best. Therefore, there are many trade-offs rather than existing an exact optimal communication system. The decision for trade-offs should consider the application requirements which diverge from one to another.} 

{
Acknowledging the diversity of UAV applications, application analysis is provided using the example of the project Drones4Energy. The application is broken down \textcolor{black}{into} essential communication categories with two data types.
}

{Concerning the} existing wireless technologies, we are mainly interested in practical measurements with UAV. It is observed that IEEE 802.11 and IEEE 802.15.4 are technologies widely used in current consumer and industrial UAV systems for remote control. The IEEE 802.11 series have advantages on throughput, comparing to IEEE 802.15.4. 
{The IEEE 802.11s amendment extends WiFi with mesh networking capabilities to provide an attractive alternative to low-power radio mesh networks.}
\textcolor{black}{LTE-based solutions grow} quickly on UAV applications, especially for \textcolor{black}{long-range} and BVLOS applications. To compensate for the capability of \ac{D2D} communication, the new protocol LTE Direct \textcolor{black}{seems} promising.
{In addition, 5G looks promising to improve the transmission latency.} LoRa is another technology recently entering into UAV applications, due to its features on \textcolor{black}{long-range}, low-cost and flexibility. It shows the possibility to maintain a \ac{UAV} coordination channel but has \textcolor{black}{limitations} on data rate and duty cycle to support \textcolor{black}{the} real-time mission data exchange such as high-resolution inspection images. 

{Adding mesh network \textcolor{black}{functionalities} to a multi-UAV system requires a new set of protocols to be deployed to obtain the needed self-organizing and self-configuration properties. The protocols should handle neighbor discovery, management of peering connections, and dynamic routing. However, only limited knowledge has been found and further research is needed for the communication protocols for UAV mesh networks.}

{The structure and interfaces for the single-master ROS network, multi-master ROS network and ROS2 network are analyzed. The node discovery procedure and transport procedure are compared between ROS and ROS2 communication middlewares. A survey of performance evaluation of robotic communication middleware is presented. It is observed that the architecture of ROS2 provides more robust network connectivity and better scalability on network size and extendability on custom plugins to default communication middleware by using DDS middleware. However, it pays extra time cost to convert the application message to \textcolor{black}{satisfy} DDS standards. On the other hand, many third-party projects are developed to improve scalability and enable real-time communication for \textcolor{black}{a} ROS network.}



%
\section*{Acknowledgment}
\textcolor{black}{This work has received funding   from the European Union’s Horizon 2020 research and innovation program under grant agreement No 861111 and the Innovation Fund Denmark project Drones4Energy with project number J. nr. 8057-00038A.}

\bibliographystyle{ieeetr}
\bibliography{references.bib}
\end{document}